\documentclass[aps,twocolumn,pra,tightenlines,floatfix]{revtex4}
\usepackage[dvips]{graphicx}
\usepackage[english]{babel}
\usepackage{amsmath}
\usepackage{amssymb}
\usepackage{times}
\newcommand{\bea}{\begin{eqnarray}}
\newcommand{\eea}{\end{eqnarray}}
\newcommand{\be}{\begin{equation}}
\newcommand{\ee}{\end{equation}}
\newcommand{\ba}{\begin{eqnarray}}
\newcommand{\ea}{\end{eqnarray}}
\newcommand{\ek}{\epsilon_{\mathbf{k}}}
\newcommand{\Ek}{E_{\mathbf{k}}}

\newcommand{\createa}[1]{a^\dagger_{#1}}
\newcommand{\destroya}[1]{a^{\phantom \dagger}_{#1}}
\newcommand{\createb}[1]{b^\dagger_{#1}}
\newcommand{\destroyb}[1]{b^{\phantom\dagger}_{#1}}

\renewcommand{\d}{\mathrm{d}}

\newcommand{\p}{\partial}

\newcommand{\phik}{\varphi_{\mathbf{k}}}

\newcommand{\xik}{\xi_{\mathbf{k}}}

\begin{document}

\title{Fermionic superfluidity: From high Tc superconductors to
  ultracold Fermi gases}

\author{Qijin Chen, Chih-Chun Chien, Yan He, and K. Levin}

\affiliation{James Franck Institute and Department of Physics,
 University of Chicago, Chicago, Illinois 60637, USA}

\date{\today}

\begin{abstract}
  We present a pairing fluctuation theory which self-consistently
  incorporates finite momentum pair excitations in the context of
  BCS--Bose-Einstein condensation (BEC) crossover, and we apply this
  theory to high $T_c$ superconductors and ultracold Fermi gases.  There
  are strong similarities between Fermi gases in the unitary regime and
  high Tc superconductors. Here we address key issues of common
  interest, especially the pseudogap. In the Fermi gases we summarize
  recent experiments including various phase diagrams (with and without
  population imbalance), as well as evidence for a pseudogap in
  thermodynamic and other experiments.
\end{abstract}

\maketitle

Superfluidity in ultracold trapped fermionic gases has been one of the
most exciting, rapidly evolving research subjects in the past few years.
\cite{Jin3,Grimma,Ketterle3a,Thomas,Salomon3a,Hulet4a}. It has captured
the attention of both condensed matter and atomic physics community, as
well as other fields such as nuclear and quark matter \cite{Wilczek}.
Atomic Fermi gases have strong similarities to superconductors.
Breakthroughs in cooling techniques in recent years have made it
possible to cool an atomic Fermi gas to quantum degeneracy temperature
and thus form a superfluid.  What is remarkable about the atomic Fermi
gases is their tunability and controllability. Using a Feshbach
resonance, one can tune the attractive two-body interaction from weak to
strong, and thereby make a smooth crossover from a BCS type of
superfluid to a Bose-Einstein condensation (BEC) \cite{Leggett}.
BCS-BEC crossover has been of interest to the condensed matter community
for a long time, and has been argued \cite{LeggettNature} to be relevant
to high $T_c$ superconductivity.

Most of the theories that are proposed to explain high $T_c$
superconductivity are based on the pairing of electrons, as in BCS
theory. While the pairing mechanism in the cuprate superconductors is
far from being clear, there is no ambiguity about the pairing
interaction in atomic Fermi gases. The attractive pairing potential
between the fermionic atoms of different spins has been well
characterized and can be computed from first principle quantum mechanics
calculations.  Here we presume that cuprate superconductivity originates
from electron pairing, so that studying superfluidity in atomic Fermi
gases may provide insights into the high $T_c$ superconductivity or vice
versa \cite{ChenPRL98,ourreviews,LeggettNature}.

Importantly, preformed pairs are an inevitable consequence as the pairing
interaction increases. At the unitary scattering limit, the superfluid
transition temperature $T_c$ is very high in units of the noninteracting
Fermi energy $E_F$. The atomic Fermi gases often have a
relatively higher transition temperature than the cuprates. For the
latter $T_c \lesssim 0.1 E_F$ is somewhat lower, owing in part to their
low dimensionality. Except in the extreme BCS limit, preformed pairs
contribute a fermionic excitation gap, i.e., a pseudogap, even in the
normal state. Such a pseudogap is common to both high $T_c$
superconductors and atomic Fermi gases.

In the absence of a population imbalance, superfluidity is associated
with a gas of fermions with dispersion $\xik = \ek -\mu$, subjected to
an attractive interaction between two different spin states, $U_{\bf
  k,k'} = U \varphi_{\bf k}\varphi_{\bf k'}$, where $U<0$ is the
coupling strength. For the cuprates, $\ek = 2t_\parallel (2-\cos k_x -
\cos k_y)$, and for atomic Fermi gases $\ek = k^2/2m$; $\mu$ is the
fermionic chemical potential, $t_\parallel$ is the in-plane hopping
integral, and we have set the lattice constant to 1.  Here and
throughout, we assume natural units and set $\hbar=k_B=c=1$.  The
function $\varphi_{\mathbf{k}}$ reflects the pairing symmetry. For the
cuprates, $\varphi_{\mathbf{k}} = \cos k_x - \cos k_y $; for the 
Fermi gases, the interaction is short-ranged, and taken to be a
contact potential so that $\varphi_{\mathbf{k}}=1$.

The Hamiltonian is given by
\begin{eqnarray}
H&-&\mu N  =  \sum_{{\bf k}\sigma} \xi_{\bf k}
a^{\dag}_{{\bf k}\sigma} a^{\ }_{{\bf k}\sigma}
\nonumber \\
& & + \sum_{\bf k k' q} U_{\bf k, k'} 
a^{\dag}_{{\bf k}+{\bf q}/2\uparrow} 
a^{\dag}_{-{\bf k}+{\bf q}/2\downarrow} 
a^{\ }_{-{\bf k'}+{\bf q}/2\downarrow} 
a^{\ }_{{\bf k'}+{\bf q}/2\uparrow},\nonumber\\
\label{eq:Hamiltonian}
\end{eqnarray}
where $a^\dag (a)$ is the creation (annihilation) operator of the
fermions.
This picture is often referred to as a one-channel model. The physics of
a Feshbach resonance can be described by a two-channel model, which
includes an open-channel, as in the one-channel model, and a
closed-channel, which represents the effects of di-atomic molecules of a
different total spin.
The two-channel Hamiltonian is given by
\begin{eqnarray}
H&-&\mu N=\sum_{{\bf k},\sigma}\xik \createa{{\bf
    k},\sigma}\destroya{{\bf k},\sigma}+\sum_{\bf q}(\epsilon_{\bf
    q}+\nu-2 \mu)\createb{\bf q}\destroyb{\bf q}\nonumber\\ 
&+&\sum_{{\bf q},{\bf k},{\bf k'}}U({\bf k},{\bf k'})\createa{{\bf
    q}/2+{\bf k},\uparrow}\createa{{\bf q}/2-{\bf
    k},\downarrow}\destroya{{\bf q}/2-{\bf k'},\downarrow}\destroya{{\bf
    q}/2+{\bf k'},\uparrow}\nonumber\\ 
&+&\sum_{{\bf q},{\bf k}}\left(g({\bf k})\createb{{\bf q}}\destroya{{\bf
    q}/2-{\bf k},\downarrow}\destroya{{\bf q}/2+{\bf
    k},\uparrow}+\text{h.c.}\right) , 
\label{eq:Hamiltonian-2ch}
\end{eqnarray}
where $b^\dag$ ($b$) is the creation (annihilation) operator for the
closed-channel molecules, whose dispersion is given by $\epsilon_{\bf
  q}=q^2/2 M$ with $M=2m$. Here $\nu$ represents the magnetic detuning,
which is used to tune the relative energy level splitting between the
open- and closed-channels and thus change the overall effective
interaction; the interaction increases as the field decreases. The last
line in Eq.~(\ref{eq:Hamiltonian-2ch}) represents the coupling between
the two channels, with $g(\mathbf{k}) = g \phik$.

Current studies of atomic Fermi gases have concentrated on $^6$Li and
$^{40}$K. For both gases, the Feshbach resonances which are used to
enhance the pairing interaction have a very big width. As a consequence,
the closed-channel fraction becomes negligible \cite{ChenClosed,Hulet4a}
even though the closed channel makes a big contribution to the effective
pairing interaction.  For example, for $^6$Li, both experiments
\cite{Hulet4a} and our theoretical calculations \cite{ChenClosed} have
demonstrated that the closed-channel population is extremely small in
the unitary regime (of the order of $10^{-5}-10^{-4}$) and remains small
in the entire range of magnetic field (i.e., pairing strength) which is
accessed experimentally.

The net effect of this tiny closed-channel fraction is that, for many
purposes, one can use the one-channel model, Eq.~(\ref{eq:Hamiltonian}),
provided one replaces $U$ with the overall effective interaction,
$U_{eff}(Q) \equiv U + g^2 D_0 (Q),$ where $D_0(Q)\equiv 1/ [ i\Omega_n-
\epsilon_{\bf q}^{mb}-\nu + 2 \mu]$ is the bare propagator of the
closed-channel bosons.  Experimentally, the pairing interaction for
atomic Fermi gases is usually parametrized by the dimensionless product,
$1/k_Fa$, where $a$ is the $s$-wave inter-fermion scattering length, and
$k_F$ is the Fermi wavevector defined in the noninteracting limit at
$T=0$, satisfying $E_F=k_F^2/2m$.

In this way both high $T_c$ superconductors and atomic Fermi gases can
be described by the same Hamiltonian but with different pairing
symmetries and free-particle dispersions. The existence of a big
pseudogap and the short coherence length in the cuprates suggest that
they are in the intermediate regime between BCS and BEC
\cite{LeggettNature}.  This is often referred to as the ``strongly
interacting regime'' in atomic Fermi gases, and is the most complex and
interesting regime as well.

Our pairing fluctuation theory uses a $T$-matrix formalism. Instead of
providing a detailed derivation, which can be found in
Refs.~\cite{ChenPRL98,ourreviews}, here we recapitulate the key
ingredients and observations of this formalism. Pairs consist of
fermions which glue together via the attractive interaction, and the
pair propagator represents a summation of multiple particle-particle
scattering processes.  Fermions acquire self-energy via creation and
destruction of a pair and a hole. In BCS theory, such pairs would be
just the Cooper pairs in the condensate, and they contribute a
self-energy $\Sigma_{sc}(K) = -\Delta_{sc}^2 G_0(-K)\phik^2$, where
$\Delta_{sc}$ is the order parameter and $G_0(K)$ is the bare fermion
Green's function.  [Throughout, we use a four-vector notation as in
Ref.~\cite{ChenPRL98}, $K\equiv (\mathbf{k},i\omega_n)$, $\sum_K =
T\sum_n\sum_{\mathbf{k}} $, etc., and $\omega_n$ is the Matsubara
frequency.]  In general, finite momentum pairs contribute to the fermion
self-energy via $\Sigma_{pg}(K) = \sum_Q t_{pg}(Q) G_0(Q-K)\phik^2$,
where $t_{pg}(Q) = 1/[1+U_{eff}(Q)\chi(Q)]$ is the $T$-matrix (pair
propagator), with $\chi(Q) = \sum_K G_0(Q-K)G(K)$ being the pair
susceptibility and $G(K)$ the full fermion Green's function.  Below
$T_c$, $t_{pg}(Q)$ diverges at $Q=0$ so that we can approximate
$G_0(Q-K) \approx G_0(-K)$ in $\Sigma_{pg}$. As a result, we obtain
$\Sigma_{pg}(K) \approx -\Delta_{pg}^2 G_0(-K)\phik^2$, with
\begin{equation}
\Delta_{pg}^2 \equiv -\sum_Q t_{pg}(Q).
\label{eq:PG}
\end{equation}
Here the key observation is that, under a reasonable approximation, the
self-energy of the finite momentum pairs has the same form as
$\Sigma_{sc}(K)$. Since these finite momentum pairs do not have phase
coherence, their contributions lead to a pseudogap in the fermion
dispersion. The total excitation gap $\Delta$ is given by
\begin{equation}
\Delta^2=\Delta_{sc}^2 + \Delta_{pg}^2 ,
\end{equation}
and the fermionic quasiparticle dispersion becomes $\Ek = \sqrt{ \xik^2
  + \Delta^2 \phik^2} $. The fact that the gaps add in quadrature can be
understood from the observation that $\Delta_{sc}^2$ and $\Delta_{pg}^2$
are proportional to the density of the condensed and noncondensed pairs,
respectively.

The effective dispersion $\Omega_{\mathbf{q}}$ and chemical potential
$\mu_{pair}$ can be determined by a Taylor expansion of $t_{pg}^{-1}$,
which, after analytical continuation ($i\Omega_n \rightarrow \Omega +i
0^+$), can be written as
\begin{equation}
  t_{pg}^{-1}  = Z(\Omega - \Omega_{\mathbf{q}} +\mu_{pair}) ,
\end{equation}
where $Z$ is the inverse residue \cite{ourreviews}, and we have
neglected the small imaginary part \cite{ourreviews}. The pairing
instability condition, or equivalently the BEC condition for pairs,
requires
\begin{equation}
t_{pg}^{-1}(0)=0\,,
\label{eq:gap}
\end{equation}
so that we have $\mu_{pair}=0$ below $T_c$.  One can readily see that
$\Delta_{pg}\rightarrow 0$ as $T\rightarrow 0$.  Therefore, this finite
$T$ theory is consistent with the BCS-Leggett ground state
\cite{Leggett}, which has been widely assumed in the literature.

The presence of the pseudogap necessarily suppresses the superfluid
transition $T_c$. This can be seen from the relationship between
$\Delta$ and $T$, as determined by a mean-field BCS gap equation: $T$
decreases as $\Delta$ increases, and $T_c$ simply corresponds to 
the lowest $T$ at which $\Delta = \Delta_{pg}$.

Our theory can be summarized in three self-consistent equations: the gap
equation (\ref{eq:gap}), the pseudogap equation (\ref{eq:PG}), and the
number equation
\begin{equation}
n = 2 \sum_K G(K) \,.
\end{equation}
They can be used to solve for $\Delta$, $\Delta_{pg}$, $\Delta_{sc}$,
$\mu$ and $T_c$ as a function of $U$ (or $U_{eff}$) or $1/k_Fa$ and $T$.

The present theory naturally leads to the following important
observations.  (i) The fundamental statistical entities in these
superfluids are fermions. We measure the ``bosonic'' (or pair) degrees
of freedom indirectly via the fermionic gap parameter $\Delta(T)$.
(ii) As we go from BCS to BEC, pairs will begin to form at a temperature
$T^*$ above $T_c$.  This pair formation is associated with a normal
state pseudogap.
(iii) In general there will be two types of excitations in these BCS-BEC
crossover systems. Importantly in the intermediate case the excitations
consist of a mix of both fermions and bosons.

It should be pointed out that effects of pairing fluctuations were first
considered by Nozi\'eres and Schmitt-Rink \cite{NSR}. However, they only
included them in the particle number equation; the absence of these
effects in the gap equations necessarily leads to internal
inconsistencies, say, in computing the superfluid density. In contrast
to the present approach, the NSR scheme also predicts that $T_c$
approaches its BEC asymptote from above for an $s$-wave superfluid in
three-dimensional (3D) continuum space.  The NSR treatment, due to its
relative simplicity, has been used by others in the recent literature
\cite{Hollanda,Parish06}.

A self-consistent treatment of finite-momentum pair excitations is
crucial.  This is more so in atomic Fermi gases, where one cannot
directly measure the temperature except in the extreme BCS and BEC
limits. Quantitatively reasonable knowledge of temperature effects is
necessary in order to determine temperature and related physical
properties in the BCS-BEC crossover regime.

\begin{figure}
\centerline{\includegraphics[width=2.8in,clip]{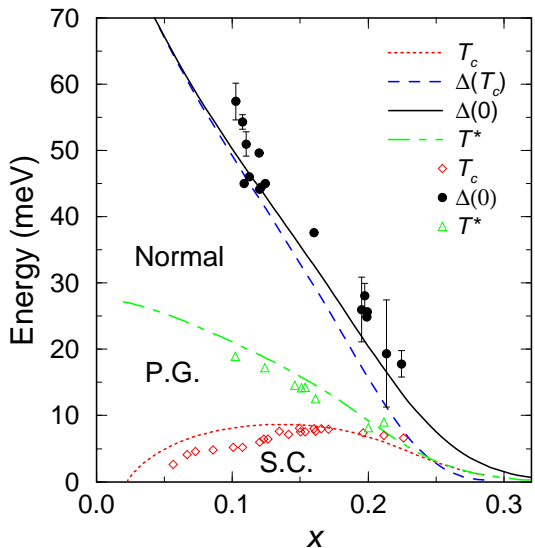}}
\caption[Cuprate phase diagram, and comparison with experiment.]
{Cuprate phase diagram showing $\Delta (0)$, $T_c$, $\Delta_{pg} (T_c)$,
  and $T^*$, calculated for $-U/4t_0=0.047$, and $t_\perp/t_\parallel
  =0.003$.  Shown as symbols are experimental data.  The normal,
  pseudogap, and superconducting phases are labeled with ``Normal'',
  ``P.G.'', and ``S.C.'', respectively. From Ref.~\cite{ChenPhD}.}
\label{fig:Cuprate_Phase}
\end{figure}

We first apply our pairing fluctuation theory to high $T_c$
superconductors. In Fig.~\ref{fig:Cuprate_Phase} we plot the calculated
phase diagram for $\mathrm{Bi_2Sr_2CaCu_2O_{8+\delta}}$ (Bi2212) and
$\mathrm{YBa_2Cu_3O_{7-\delta}}$ (YBCO) single crystals, as well as
present a comparison with experimental data.  The horizontal axis is the
hole doping concentration per unit cell, $x= 1-n$. Here we show
theoretical calculations and experimental data for zero $T$ excitation
gap $\Delta(0)$, superconducting transition temperature $T_c$, and
pairing onset temperature $T^*$. We also show the theoretical result for
$\Delta(T_c)$, the pseudogap at $T_c$.

To obtain this phase diagram, we fit with one parameter $-U/4t_0$ such
that it gives the right value of $T_c$ at optimal doping around
$x=0.15$. Here $t_0$ is the bare in-plane electron hopping matrix
element, in the absence of the Coulomb repulsion. It is related to
$t_\parallel$ via $t_\parallel \approx t_0 x$. One would obtain
essentially the same phase diagram, if one parametrized the variable
attractive interaction by simply fitting $T^*(x)$ to experiment.  Thus,
the underdoped regime sees effectively a stronger pairing interaction,
$-U/4t_\parallel$.  It is evident that our simple model gives a
(semi-)quantitatively satisfying phase diagram, in agreement with
experiment.

It should be noted that, in contrast to the $s$-wave, 3D Fermi gas, the
$d$-wave symmetry and the lattice periodicity in the cuprates makes
$T_c$ vanish before the BEC regime can be reached as the couping strength
increases \cite{Chen1}. This introduces a lower critical doping $x_c$ in
the phase diagram shown in Fig.~\ref{fig:Cuprate_Phase}.

\begin{figure}
\centerline{ \includegraphics[width=2.45in,clip]{Hc1_7.eps}}\vskip 1ex
\centerline{ \includegraphics[width=2.3in,clip]{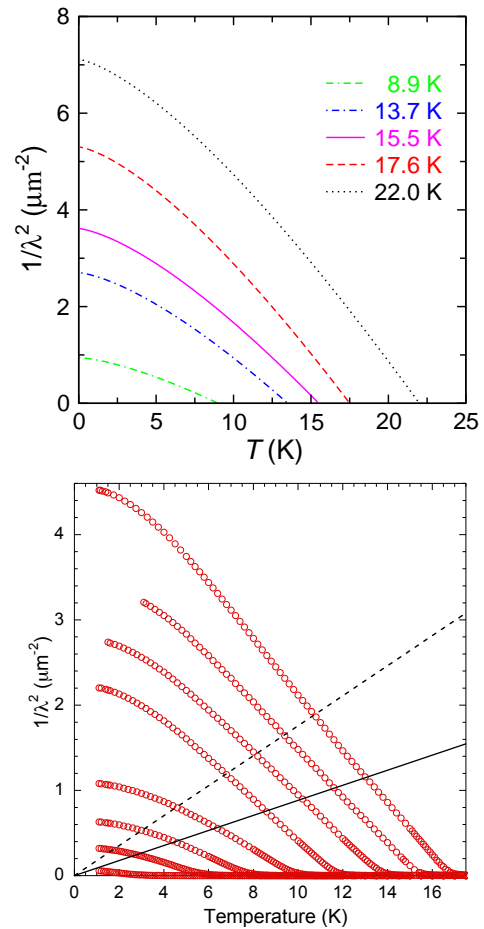}}
\caption{Comparison of superfluid density between theory and experiment.
  Top panel: Theoretical calculation for underdoped cuprate
  superconductors; Bottom panel: Experimental data from
  Ref.~\cite{Hardynew} in the extreme underdoped regime of YBCO.}
\label{fig:Ns}
\end{figure}

Other applications of BCS-BEC crossover theory to the cuprate
superconductors include studies \cite{ChenPRL98} of the quasi-universal
behavior of the normalized superfluid density $n_s(T)/n_s(0)$ as a
function of $T/T_c$ for various doping concentrations. Experimental data
from Ref.~\cite{Hardynew} are plotted in the lower panel of
Fig.~\ref{fig:Ns} for a series of very underdoped cuprates. The
universality apparent in these data presents a puzzle for conventional
theories of the penetration depth which invoke only fermionic
excitations of the condensate.  This is because the fermionic
contributions are expected to reflect the strong variations in the
excitation gap $\Delta(x)$ associated with the different hole
concentrations $x$, thereby leading to highly non-universal behavior. In
the theory plot of the upper panel (Figure \ref{fig:Ns}) bosonic or pair
excitations contribute an extra mechanism for destroying superfluidity.
They compensate for the decrease of the quasiparticle contribution
($\propto T_c(x)/\Delta(x)$) as the system evolves from overdoping to
underdoping. As a result, the slope of the superfluid density remains
nearly doping independent, with only a small systematic variation with
$x$. The comparison between theory and recent experiments from the UBC
group \cite{Hardynew} shows quite good agreement.
Recently, the bosonic nature of the superfluid density in underdoped
cuprates has also been recognized by other groups
\cite{Herbut,UemuraNs}.

\begin{figure}
\centerline{\includegraphics[width=3.3in,clip]{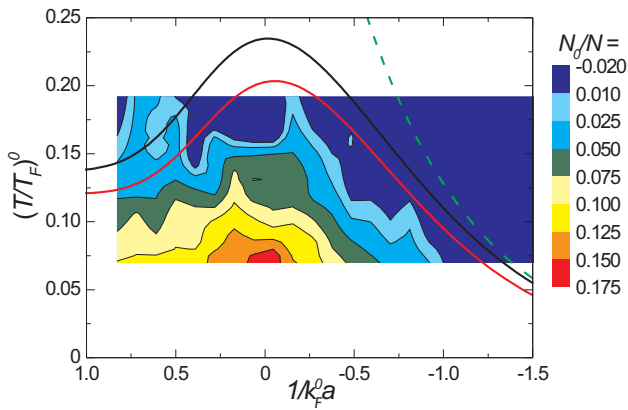}}
\caption{(color) Phase diagram of $^{40}$K. A contour plot of the
  measured condensate fraction $N_0/N$ as a function of $1/k_F^0a$ and
  effective temperature $(T/T_F)^0$ is compared with theoretically
  calculated contour lines at $N_s/N=0$ (black curve) and 0.01 (red
  curve).  The overall trend of the experimental contour of $N_0/N=0.01$
  and the theoretical line for $N_s/N=0.01$ are in good agreement.  The
  dashed line represents the naive BCS result, $T_c/T_F^0 \approx 0.615
  e^{\pi/2k_F^0a}$.  From Ref.~\cite{Jin_us}.  }
\label{fig:K40_Phase}
\end{figure}

We turn now to experiments in cold atom systems. Here we
\cite{JS5,ourreviews} need to include the trap potential $V(r)$. To this
end, we use a local density approximation (LDA) by replacing
$\mu\rightarrow \mu(r) \equiv \mu -V(r)$.  Here $\mu\equiv \mu(0)$ is
the global chemical potential and $V(r) = m\bar{\omega}^2 r^2/2 $ for a
harmonic trap with mean angular frequency $\bar{\omega}$.  The number
constraint enters for the entire trap, $N=\int \mathrm{d}^3r \,n(r)$.
Below $T_c$, there exists a superfluid core within $r<R_{sc}$. For
$r>R_{sc}$, $\mu_{pair} < 0 $, and the gap equation has to be modified
to reflect this non-zero $\mu_{pair}$: $t_{pg}^{-1}(0)= Z\mu_{pair}$.

Before we compare with experiment, we have to know the temperature of
the system.  Experimentally, temperature is measured at the BCS
\cite{Jin3} or BEC end \cite{Grimm4a}, before or after the experiment is
performed in the strongly interacting regime. This BCS or BEC state is
adiabatically connected to the state where experiments are done via a
slow magnetic field sweep. By including self-consistently finite
momentum pair excitations, our pairing fluctuation theory enables us to
calculate the entropy $S(T)$ at an arbitrary magnetic field. Then we can
map the physical temperature $T$ onto that measured in the BCS or BEC
limit, $T^0$, or vice versa \cite{ChenThermo}.

In Fig.~\ref{fig:K40_Phase}, we show the phase diagram for $^{40}$K and
compare our calculations with experimental data \cite{Jin_us}. Here all
temperatures are measured in (or converted into) the free Fermi gas
limit.  In addition, $T_F^0\equiv k_F^0/2m$ is the Fermi temperature at
the trap center in the noninteracting limit.  The experimental
superfluid phase boundary is given by the 1\% condensate fraction
contour line; it is hard to locate experimentally where the condensate
vanishes precisely.  This should be compared with the (red) theory
curve.  Here we \cite{Jin_us} identify the condensate fraction with the
superfluid density.  The agreement between experiment and theory is
reasonably good. It should be noted that both experiment and theory show
a maximum around unitarity, $1/k_F^0a = 0$, which has also been observed
in $^6$Li phase diagrams \cite{Ketterle3a}.

\begin{figure}
\centerline{\includegraphics[width=3.0in,clip]{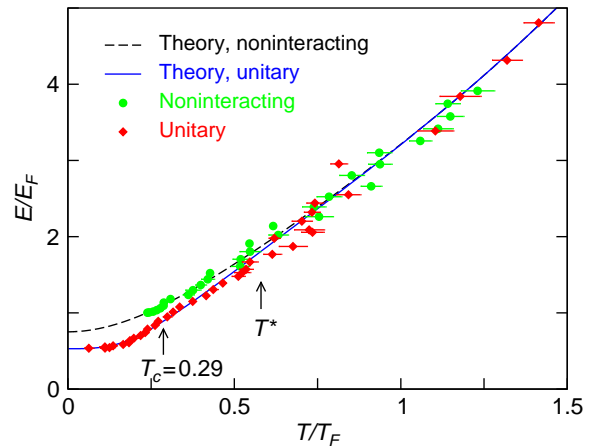}}
\caption{(color) Energy per particle as a function of $T$ for $^6$Li at
  unitarity.  The fact that the experimental data (symbols) (and the two
  theoretical curves) for noninteracting and unitary Fermi gases do not
  merge until higher $T^* > T_c$ is consistent with the presence of a
  normal state pseudogap. From Ref.~\cite{ThermoScience2}. }
\label{fig:EvsT}
\end{figure}

We next turn to a detailed comparison of theory and experiment for
thermodynamics.  Figure \ref{fig:EvsT} presents a plot of energy $E$ as
a function of $T$ comparing the unitary and non-interacting regimes for
$^6$Li.  The solid curves are theoretical while the symbols are
experimental data \cite{ThermoScience2}.  A realistic Gaussian trap with
the experimentally given trap depth was used. There has been a
recalibration of the experimental temperature scale
\cite{ThermoScience2} in order to plot theory and experiment in the same
figure.  The latter was determined via Thomas-Fermi fits to the density
profiles \cite{JS5}.  To arrive at the calibration, we applied the same
fits to the theoretically produced density profiles.

Good agreement between theory and experiment is apparent in
Fig.~\ref{fig:EvsT}. It should be emphasized that there is \emph{no}
fitting parameter in our calculations.  The temperature dependence of
$E$ reflects primarily fermionic excitations at the edge of the trap
\cite{ChenThermo}, although there is a small bosonic contribution as
well.
Importantly one can see the effect of a pseudogap in the unitary case.
That is, the temperature $T^*$ is visible from the plots as that at
which the non-interacting and unitary curves merge.  This corresponds
roughly to $T^* \approx 2 T_c$.  Evidence of a pseudogap at unitarity
has also been observed more directly in a radio-frequency excitation
gap experiment \cite{Grimm4a}.

Finally, we turn to the effects of population imbalance, which have
recently been one of the hottest subjects in atomic Fermi gases. We
consider both the homogeneous system \cite{Chien06}, as well as trapped
case, which can be found in Refs.~\cite{ChienRapid,Chien07}.  We define
$n = n_\uparrow + n_\downarrow$, $\delta n = n_\uparrow - n_\downarrow$,
and the polarization $p = \delta n /n $.  There will be two number
equations associated with the two spin species. The polarized superfluid
phase is conventionally called the ``Sarma'' phase \cite{Sarma63},
although here it is generalized away from the BCS limit originally
considered by Sarma. Details are given in Ref.~\cite{Chien06}.

\begin{figure}[t]
  \centerline{\includegraphics[width=3.2in,clip]{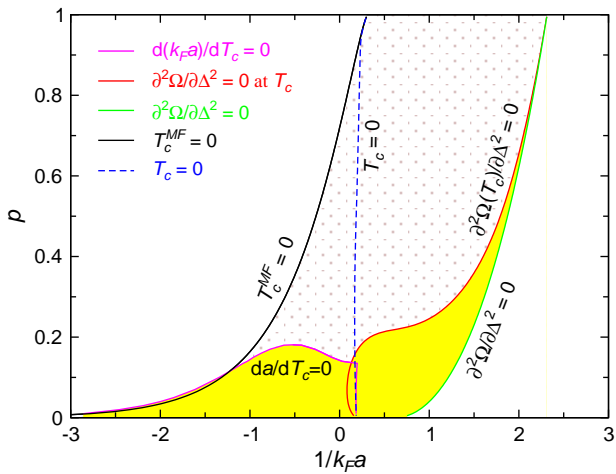}}
  \caption{(color) Phase diagram in the $p-1/k_Fa$ plane, showing where
    intermediate temperature superfluidity associated with the Sarma
    phase (shaded region) exists.  The (red) curve labeled by $\p^2
    \Omega(T_c)/\p\Delta^2=0$ is determined at corresponding $T_c$. The
    line defined by $\d a/\d T_c =0$ is given by the turning points
    $(p,1/k_Fa)$ where $1/k_Fa$ reaches a local extremum as a function
    of $T_c$.  At $T=0$, the entire region between the $T_c^{MF}=0$ and
    the $\p^2 \Omega/\p\Delta^2=0$ lines is unstable against phase
    separation. However, a A stable polarized, intermediate temperature
    superfluid phase exists for the (yellow) shaded region.  Homogeneous
    superfluids in the dotted region are unstable at any temperature.
    From Ref.~\cite{ChenStability}.}
\label{fig:Imb_Phase}
\end{figure}

In Fig.~\ref{fig:Imb_Phase}, we present the zero $T$ phase diagram for a
population imbalanced homogeneous Fermi gas in the $p-1/k_Fa$ plane.
Here $T_c^{MF}$ is the solution of $T_c$ in a strict mean-field
treatment.  $\Omega$ is the thermodynamic potential. A superfluid
solution is stable only if $\partial^2 \Omega/\partial\Delta^2 > 0 $.
From this figure, we see that at any $p\ne 0$, there is a minimum
threshold for $1/k_Fa$ (determined by $T_c^{MF}=0$) in order to have a
superfluid solution.  The point at which the solution becomes stable
(determined by $\partial^2 \Omega/\partial\Delta^2 =0$) occurs in BEC
regime. The (blue) dashed line indicates where $T_c$ vanishes. The
dotted region (as well as the shaded region) in the figure indicates an
unstable (Sarma) superfluid phase at $T=0$. The stable phases in this
region are presumed to be either phase separated (PS) or possibly a
Fulde-Ferrell-Larkin-Ovchinnikov state \cite{LOFF1}.

In addition to the zero $T$ phases, we also show in
Fig.~\ref{fig:Imb_Phase} where the system is stable at $T_c$. A stable
finite temperature polarized superfluid exists in the (yellow) shaded
region. This is a very interesting prediction of our theory which
appears consistent with experiment. More details can be found in
Ref.~\cite{Chien06}.

\begin{figure}[b]
\centerline{\includegraphics[clip,width=3.2in] {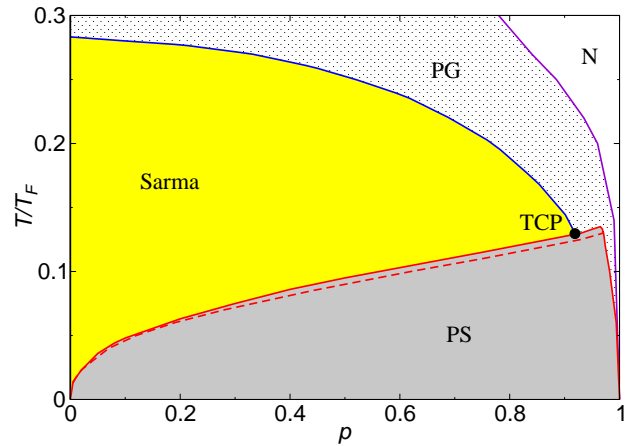}}
\caption{(color) Phase diagram of a population-imbalanced Fermi
gas in a harmonic trap at unitarity. The solid lines separate
different phases. Above the (red) dashed line but within the PS phase,
the superfluid core does not touch the domain wall.
Here ``PG'' indicates the pseudogapped normal phase.  The black dot
labeled ``TCP'' indicates the tricritical point. 
}
\label{fig:TP_trap}
\end{figure}

Finally, we show in Fig.~\ref{fig:TP_trap} the phase diagram at
unitarity for a population-imbalanced Fermi gas in a harmonic trap with
angular frequency $\omega$.  Here $1/k_{F}a=0$.
Phase separation (labeled PS) occupies the lower $T$ portion of the
phase diagram, where the gap $\Delta$ jumps abruptly to zero at some
trap radius.  At intermediate $T$, there is a (yellow-shaded) Sarma
phase, where $\Delta$ vanishes continuously within the trap.  It evolves
into a (dotted region) pseudogap (PG) phase as the superfluid core
vanishes at higher $T$.  A normal (N) phase without pairing always
exists at even higher $T$.  This phase diagram appears to be consistent
with current experiments. One can easily see that the physics associated
with population imbalance is much richer than in the absence
of imbalance.

In conclusion, in this paper we have addressed commonalities
particularly associated with the phase diagrams of ultracold trapped
fermionic gases and high $T_c$ superconductors.  These common features
revolve around the scenario \cite{LeggettNature,ourreviews} that the
cuprates are somewhere intermediate between BCS and BEC. Importantly,
this scenario has been directly realized in trapped Fermi gases. Here
one sees considerable evidence for pre-formed pairs and the related
fermionic pseudogap which appear reminiscent of their cuprate
counterparts.

This work was supported by NSF PHY-0555325 and NSF-MRSEC Grant
No.~DMR-0213745.

\bibliographystyle{apsrev}


\end{document}